\documentclass{sig-alternate-05-2015}
\usepackage{hyperref}
\usepackage{bm}
\usepackage{url}
\usepackage{xspace}
\usepackage{amsmath}
\usepackage{amssymb}
\usepackage{color}
\usepackage{algorithm}
\usepackage{algorithmic}
\usepackage{multirow}
\usepackage{booktabs}
\usepackage{times}
\usepackage{breqn}
\usepackage{graphicx} 
\usepackage{subfigure} 
\newcommand{\cfnade}{CF-NADE\xspace}

\newcommand{\reals}{\mathbb{R}}

\newcommand{\cost}{\mathcal{C}}

\newboolean{for_submission}
\setboolean{for_submission}{false}
\newcommand{\says}[3]{\ifthenelse{\boolean{for_submission}}{}{{\color{#3}#1 says: \emph{#2}\color{black}}\xspace}}

\begin{document}

\setcopyright{acmcopyright}





%

\title{Neural Autoregressive Collaborative Filtering \\for Implicit Feedback}
%
%
%
%
%

\numberofauthors{4} 
%
\author{
%
%
\alignauthor
Yin Zheng\\
       \affaddr{Hulu LLC.}\\
       \affaddr{Beijing, China, 100084}\\
       \email{yin.zheng@hulu.com}
\alignauthor
Cailiang Liu\\
       \affaddr{Hulu LLC.}\\
       \affaddr{Beijing, China, 100084}\\
       \email{cailiang@hulu.com}
\alignauthor Bangsheng Tang\\
       \affaddr{Hulu LLC.}\\
       \affaddr{Beijing, China, 100084}\\
       \email{bangsheng@hulu.com}
\and  
\alignauthor Hanning Zhou\\
       \affaddr{Hulu LLC.}\\
       \affaddr{Beijing, China, 100084}\\
       \email{eric.zhou@hulu.com}
}
\CopyrightYear{2016} 
\setcopyright{acmlicensed}
\conferenceinfo{DLRS '16,}{September 15 2016, Boston, MA, USA}
\isbn{978-1-4503-4795-2/16/09}\acmPrice{\$15.00}
\doi{http://dx.doi.org/10.1145/2988450.2988453}
\maketitle
\begin{abstract}
  
  This paper proposes \emph{implicit} \cfnade, a neural autoregressive
  model for collaborative filtering tasks using implicit feedback(
  e.g. click/watch/browse behaviors). We first convert a user's
  implicit feedback into a ``like'' vector and a confidence vector, and
  then model the probability of the ``like'' vector, weighted by the
  confidence vector. The training objective of implicit \cfnade is to
  maximize a weighted negative log-likelihood. We test the performance
  of implicit \cfnade on a dataset collected from a popular digital TV
  streaming service. More specifically, in the experiments, we
  describe how to convert watch counts into implicit ``relative
  rating'', and feed into implicit \cfnade. Then we compare the
  performance of implicit \cfnade model with the popular implicit
  matrix factorization approach. Experimental results show that
  implicit \cfnade significantly outperforms the baseline.

\end{abstract}

%
%


%
%

%
%




\section{Introduction}
\label{sec:intro}
Modern online systems rely heavily on recommender systems to help
users identify items they might be interested in, from a usually
massive catalog, and therefore provide personalized experience. The
most popular and successful technique of building a recommender system
is \emph{collaborative filtering} (CF) \cite{goldberg1992using}, which
predicts user preferences by analyzing past user behaviors and
establishing relevance between items and also between users. There are
two major types of inputs to a CF-based recommender system: 1)
\emph{explicit feedback}, e.g. 5-star ratings, likes/dislikes; and 2)
\emph{implicit feedback} e.g. watch/search/browse/purchase
behaviors. Explicit feedback accurately reflects a user's preference
over an item, and thus is most convenient to use. Techniques designed
for explicit feedback, such as restricted Boltzmann machine (RBM) CF
\cite{salakhutdinov2007restricted}, matrix factorization
\cite{mnih2007probabilistic,salakhutdinov2008bayesian,koren2009matrix},
neural network matrix factorization\cite{dziugaite2015neural}, and
recently developed \emph{neural autoregressive distribution estimator
  for CF tasks} (\cfnade) \cite{zheng2016neural} have been highly
successful in predicting explicit user preferences, which, to the best
of our knowledge, is the state-of-the-art on MovieLens 1M, MovieLens
10M~\cite{harper2015movielens} and Netflix
datasets~\cite{bennett2007netflix}.

In real-world applications, only a small fraction of users actively
provide explicit feedback, which restricts the application of
aforementioned methods. On the other hand, implicit feedback is
abundant, as long as the user interacts with the online system. Hence,
building recommender system using collaborative filtering on implicit
feedback has attracted increasing attention. One major characteristic
of implicit feedback is that there is only \emph{positive} feedback,
in that one can only tell whether a user \emph{has} engaged with an
item for how many times. Consider the size of the catalog, the number
of items a user has engaged with is tiny. So implicit feedback is
inherently \emph{unbalanced} and \emph{sparse}. Also, a user has not
engaged with an item does not necessarily mean that he/she dislikes
the item or the item is irrelevant, and in most cases it is because
the user is unaware of the item. Therefore in literature,
collaborative filtering using implicit feedback is sometimes referred
to as \emph{one-class collaborative filtering} (OCCF)\cite{pan2008one}.

A natural way of building recommender system using collaborative
filtering for implicit feedback is to interpret implicit feedback as
explicit feedback in a proper way and apply existing successful
algorithms for explicit feedback, such as
\cite{hu2008collaborative,pan2008one}.  In this paper, we describe a
generalized \cfnade for implicit feedback, which is referred to as
\textit{implicit} \cfnade.  Specifically, we first introduce the
original \cfnade model~\cite{zheng2016neural} for explicit feedback
briefly in Section~\ref{sec:cfnade}. Then, we focus on describing
implicit \cfnade in Section~\ref{sec:implicit_cfnade}. We compare
implicit \cfnade with Implicit Matrix Factorization (IMF)
approach~\cite{hu2008collaborative}, and show the performance
comparison in Section~\ref{sec:exp}.


\section{Related Work}
Many previous works on recommender system using implicit feedback are
based on matrix factorization. \cite{hu2008collaborative} proposes to
employ matrix factorization where implicit feedback is treated as
binary preferences and weighted according to the number of
engagements. The work of \cite{hu2008collaborative} is quite popular
and has been included into popular software libraries like MLlib of
Spark~\cite{meng2015mllib}. \cite{pan2008one} also formulates the
problem as weighted matrix factorization, and proposes to use negative
sampling to mitigate the unbalancedness problem. With a similar
weighting strategy, inner products are replaced by logistic functions
in the probabilistic model called logistic matrix factorization
\cite{johnson2014logistic}. Besides weighting, values can be imputed
for unobserved examples to indicate possible feedback. This method and
its combination with weighting are discussed in \cite{yao2014dual}.  In
\cite{li2010improving}, multiple implicit feedback sources are
considered, either by treating each source separately and combining
with a linear model, or collectively embedding all feedback sources
into a collective collaborative filtering model. Alternatively, SLIM
\cite{ning2011slim} formulates the CF for implicit feedback as a
convex optimization problem, which is recently generalized to LRec in
\cite{AAAI1612333}.

With the recent success of deep learning in computer vision and
natural language processing
community~\cite{krizhevsky2012imagenet,szegedy2014going,he2015deep,
  mikolov2013distributed}, neural networks have also found application
in building recommender systems. For example,
RBM-CF~\cite{salakhutdinov2007restricted} and
AutoRec~\cite{sedhain2015autorec} are successful approaches to model
the users' explicit feedback, using restricted Boltzmann machine and
autoencoder respectively. The recently developed
\cfnade~\cite{zheng2016neural} models explicit feedback with a neural
autoregressive architecture.
In this work, we will generalize \cfnade and propose a novel CF model
for implicit feedback.

\section{\cfnade}
\label{sec:cfnade}
We start with the description of \cfnade, a neural autoregressive
architecture for CF tasks which has proved successful in modeling
\textit{explicit} ratings~\cite{zheng2016neural}. A user $u$'s
explicit ratings are denoted as
$\mathbf{r}^{u} = (r^{u}_{m_{o_1}}, r^{u}_{m_{o_2}}, \ldots,
r^{u}_{m_{o_D}})$,
where $D$ is the number of items that the user has rated,
$m_i\in \{1,2,\ldots,M\}$ is the index of the $i$th rated items, $M$
is the total number of items, $o$ is a $D$-tuple in the set of
permutations of $(1,2,\ldots,D)$ which serves as an ordering of the
$D$ rated items, and $r^{u}_{m_{o_i}} \in \{1,2,\ldots,K\}$ denotes
the rating that the user gave to item $m_{o_i}$. For simplicity, we
will omit the index $u$ of ${\bf r}^u$. As discussed in
\cite{zheng2016neural}, a random order of the ratings works well in
practice and is the key to extend \cfnade to a deep model.

By the chain rule, \cfnade models the joint probability of the rating
vector $\bf r$ as a product of conditionals:
\begin{equation}
    p\left({\bf r}\right) = \prod_{i=1}^{D} p\left(r_{m_{o_i}}|{\bf r}_{m_{o_{<i}}}\right)
    \label{eqn:chain_rule}
\end{equation}
where ${\bf r}_{m_{o_{<i}}} = (r_{m_{o_1}}, r_{m_{o_2}}, \ldots, r_{m_{o_{i-1}}})$ denotes the first $i-1$ elements of $\bf r$ indexed by $o$.

Each conditional in Equation~\ref{eqn:chain_rule} is modeled as:
\begin{eqnarray}
p\left ( r_{m_{o_i}}=k|{\bf r}_{m_{o_{<i}}} \right ) &=& \frac{\exp \left(s_{m_{o_i}}^{k}\left({\bf r}_{m_{o_{<i}}}\right)\right)}{\sum_{k'=1}^K\exp \left(s_{m_{o_i}}^{k'}\left({\bf r}_{m_{o_{<i}}}\right)\right)}
\label{eqn:cfnade_softmax} 
\end{eqnarray}
$s_{m_{o_i}}^{k}({\bf r}_{m_{o_{<i}}})$ is the score indicating the
preference that the user gave rating $k$ for item $m_{o_i}$ given
previous ratings ${\bf r}_{m_{o_{<i}}}$. The score of
$s_{m_{o_i}}^{k}({\bf r}_{m_{o_{<i}}})$ is computed by:
\begin{equation}
  s_{m_{o_i}}^{k}\left({\bf r}_{m_{o_{<i}}}\right) =d^{k}_{m_{o_i}} +\mathbf{V}^{k}_{m_{o_i},:}\mathbf{h}\left ( {\bf r}_{m_{o_{<i}}} \right )
\label{eqn:score}
\end{equation}
where ${\bf V}^{k}\in \reals^{M \times H}$ and
${\bf d}^{k} \in \reals^{M}$ are the connection matrix and the bias
with rating $k$, respectively. And
\begin{eqnarray}
\mathbf{h}\left ( {\bf r}_{m_{o_{<i}}} \right ) &=& {\bf g}\left( \mathbf{b}+\sum_{j<i}\mathbf{W}^{r_{m_{o_j}}}_{:,m_{o_j}} \right ) 
\label{eqn:cfnade_hidden}
\end{eqnarray}
where ${\bf W}^k\in \reals^{H\times M }$ is the connection matrix
associated with rating $k$, ${\bf W}_{:,j}^k\in \reals^{H}$ is the
$j$\textsuperscript{th} column of ${\bf W}^k$ and $W^{k}_{i,j}$ is an
interaction parameter between the $i$\textsuperscript{th} hidden unit
and item $j$ with rating $k$, ${\bf b}\in \reals^H$ is the bias term,
$\bf g(\cdot)$ is the activation function, such as
$\tanh(x) = \frac{\exp(x)-\exp(-x)}{\exp(x)+\exp(-x)}$,

Fitting \cfnade can be simply maximizing the joint probability
$p({\bf r})$. As noticed in \cite{zheng2016neural}, maximizing the
conditional of Equation~\ref{eqn:cfnade_softmax} can only ensure that
the probability of the true rating is the largest among all possibles,
while leaving the ordinal nature of ratings disregarded. Hence, a
ranking loss is proposed to be added in the objective function, and a
significant improvement can be observed.


\section{Implicit \cfnade}
\label{sec:implicit_cfnade}
As discussed in Section~\ref{sec:intro}, implicit feedback is abundant
and easy to obtain. In this section, we describe how to adapt \cfnade
to \textit{implicit} feedback. In the implicit feedback scenario, the
``rating'' $r^u_{i}\in \reals_{\geq 0}$ that a user $u$ gives to an item
$i$ is defined as the number of times that the user interacts with the
item. Inspired by \cite{hu2008collaborative}, we could define a binary
scalar $t^u_{i}$ by binarizing the $r^u_{i}$ values:
\begin{equation}
t^u_{i} = \begin{cases}1 & r^u_{i} > 0\\0 & r^u_{i}=0\end{cases}
\label{eqn:t}
\end{equation}
where $t^u_{i}=1$ indicates user $u$ likes item $i$ as he/she has
interacted with this item before, and if user $u$ never interacted
with item $i$, we think that there is no preference of user $u$ on
item $i$. However, deciding whether a user likes or dislikes an item
by binarizing $r^u_{i}$ is oversimplified and can be quite noisy. The
reason is twofold: 1) in most cases, a user has not interacted with an
item is because of unawareness, not dislike; and 2) the number of
times the user interacts with the item can be a good indicator of how
much the user likes the item, which is lost in $\bf t$ after the
binarization.

Hence, we need to formalize the confidence that a user likes or
dislike an item.  Generally speaking, the confidence $c^u_{i}$ should
increase with $r^u_{i}$.
In this work, we follow \cite{hu2008collaborative} and define the
confidence as:
\begin{equation}
c^u_{i} = 1+\alpha r^u_{i}
\label{eqn:confidence}
\end{equation}
where $\alpha$ is the rate of confidence, which is a hyper-parameter
controlling how fast the confidence $c^u_{i}$ increases with
$r^u_{i}$. We will show the impact of $\alpha$ in the experiments.

A user $u$'s implicit feedback can now be represented as
${\bf t}^u=(t^u_{1}, t^u_{2}, \ldots, t^u_{M})$, with corresponding
confidence levels as ${\bf c}^u=(c^u_{1}, c^u_{2}, \ldots, c^u_{M})$.
In the rest of the paper, we will omit the superscript $u$ for
simplicity. Similar to \cfnade, we model the probability of
${\bf t}^u$ as a product of conditionals, with the addition of
confidence levels in the condition as:
\begin{equation}
p\left({\bf t}| {\bf c}\right) = \prod_{i=1}^{M} p\left( t_i|{\bf t}_{<i}, {\bf c}\right)
\label{eqn:implicit_chain_rule}
\end{equation}
where $M$ is the number of items, and ${\bf t}_{<i}$ denotes
preference of the previous $i-1$ items. Similar to \cfnade model, the
order of the items in Equation~\ref{eqn:implicit_chain_rule} is
randomly shuffled. As the confidence level $c_i$ should be paired with
$t_i$, Equation~\ref{eqn:implicit_chain_rule} can then be rewritten
as:
\begin{equation}
p\left({\bf t}| {\bf c}\right) = \prod_{i=1}^{M} p\left( t_i|{\bf t}_{<i}, {\bf c}_{<i}\right)
\label{eqn:implicit_chain_rule_2}
\end{equation}


To define the conditionals in
Equation~\ref{eqn:implicit_chain_rule_2}, we first define the hidden
representation given previous $i-1$ ratings as:
\begin{equation}
\mathbf{h}\left ({\bf t}_{<i}, {\bf c}_{<i} \right ) = {\bf g}\left( \mathbf{b}+{\bf W}_{:,<i}({\bf t}_{<i}\odot {\bf c}_{<i})+ {\bf A}_{:,<i}(((1-{\bf t}_{<i})\odot {\bf c}_{<i})) \right ) 
\end{equation} 
where $\bf g(\cdot)$ is the activation function, $\odot$ is
element-wise product, ${\bf W}\in \reals^{H\times M}$ and
${\bf A}\in \reals^{H\times M}$ are connection matrices associated
with the \textit{``like''} vector ${\bf t} = (t_1, t_2, \ldots, t_M)$
and the \textit{``dislike''} vector $1-{\bf t}$,
$\mathbf{b}\in \reals^H$ is the bias vector, ${\bf X}_{:,<i}$ is the
first $i-1$ columns of matrix $\bf X$, and similarly, ${\bf x}_{<i}$
is the first $i-1$ elements of vector $\bf x$, and $\bf g$ is the
activation function.  Note that if there is no confidence vector
$\bf c$, the \textit{``dislike''} connection matrix $\bf A$ would be
redundant, as the difference between the corresponding parameters
$W_{:,i}$ and $A_{:,i}$ for item $i$ would be constant, in which case
we can set $\hat{W}_{:,i}=W_{:,i}-A_{:,i}$ and $A_{:,i}=0$. However,
as the confidence varies for each user, $\bf A$ is required to capture
the differences induced by varying confidence levels.

Then the conditionals in Equation~\ref{eqn:implicit_chain_rule_2} can be modeled as:
\begin{equation}
p\left ( t_i=1|{\bf t}_{<i}, {\bf c}_{<i} \right ) = \textup{sigm}\left(d_i + \mathbf{V}_{i,:}\mathbf{h}\left(\mathbf{t}_{<i},\mathbf{c}_{<i}\right)\right)
\end{equation}
where $\mathbf{V}\in \reals^{M\times H}$ and
$\mathbf{d}\in \reals^{M}$ are the connection matrix and bias, and
$\mathbf{V}_{i,:}$ and $\mathbf{d}_{i}$ are the corresponding $i$th
row and element, respectively. $\textup{sigm}(x)$ denotes the sigmoid
function: $\frac{1}{1+\exp(-x)}$.

Training an implicit \cfnade can be done by minimizing the negative
log-likelihood of Equation~\ref{eqn:implicit_chain_rule_2} directly,
as the original \cfnade. However, due to the noisiness of $\bf t$, it
would be beneficial to incorporate confidence levels $\bf c$ to
reflect the uncertainty of the elements in $\bf t$. Hence, we
formalize the cost function as:
\begin{equation}
\cost = -\sum_{i=1}^M c_i\log p\left(t_i| {\bf t}_{<i}, {\bf c}_{<i}\right).
\label{eqn:weighted_cost}
\end{equation}

In this way, fitting an element
$p\left(t_i| {\bf t}_{<i}, {\bf c}_{<i}\right)$ wrong with high
confidence in Equation~\ref{eqn:weighted_cost} would cost more than
one with low confidence. After the model is trained, we could predict
the users' preference for each item as:
\begin{eqnarray}
p(t_i = 1| {\bf t}, {\bf c}) &=& \textup{sigm}\left(d_i + \mathbf{V}_{i,:}\mathbf{h}\left(\mathbf{t},\mathbf{c}\right)\right)\label{eqn:test_cond}\\
\mathbf{h}\left(\mathbf{t},\mathbf{c}\right) &=&  {\bf g}\left( \mathbf{b}+\left({\bf W}({\bf t}\odot {\bf c})+ {\bf A}((1-{\bf t}\right)\odot {\bf c}))\right ) 
\end{eqnarray}

Thus, the model can learn to balance the huge number of low confident
items, which are unobserved or interacted few times by the user, and
the items with high confidences but of small quantity. As a result,
the model could predict a user's preferences on unobserved items even
if the we set $t_i=0$ (dislike) as the input. And it might also happen
that a user has interacted with an item $i$ before ($t_i=1$) but for
only a few time, and the model predict that he/she does not like the
item much. For example, a user watches the first episode of a TV show,
but he/she never starts the next episode. In this situation, the
corresponding $t_i$ is $1$ and $c_i$ is small. Then the model can
learn to predict a higher probability for
$p(t_i = 0| {\bf t}, {\bf c})$.

As is noticed by~\cite{Uria2013b,zheng15deep, zheng2016neural}, with a
randomly sampled ordering (c.f. Section~\ref{sec:cfnade}), minimizing
the negative log-likelihood for NADE based model is equivalent to
randomly splitting the input vector into two parts, and treating the
first part as the input, and optimizing to maximize the conditionals
of the elements in the other part. We refer the reader to
\cite{zheng2016neural} for more details.

Thus, the training objective can be rewritten as:
\begin{equation}
\cost = \frac{M}{M-i+1}\sum_{j\ge i} -c_i\log p\left(t_{o_i}|{\bf t}_{o_{<i}}, {\bf c}_{o_{<i}}\right).
\label{eqn:deep_cost}
\end{equation}
where $o$ is a random ordering of all items that are sampled at each
training update, and $o_i$ and $o_{<i}$ are the $i$th element and
first $i-1$ elements of $o$. In practice, we will adopt this training
objective as it is easy to implement and can be extended to a deep
version efficiently.

\section{Experiments}
\label{sec:exp}

In this section, we test the performance of implicit \cfnade on a
dataset extracted from a digital TV streaming service, and compare the
performance of implicit \cfnade with the widely used Implicit Matrix
Factorization approach~\cite{hu2008collaborative}, as implemented in
Spark MLlib~\cite{meng2015mllib}.

\subsection{Dataset Description}
\label{sec:data}
Our dataset is composed of watch behaviors of $444480$ randomly
sampled active users on $17348$ movies or TV shows. Here an
\emph{item} is either a movie or a TV show. For each user, we track
the number of times that the user ``completely'' watched a movie or an
episode of a TV Show during a period of 3 years from April 2013 to
April 2016. We define a ``complete'' watch as any continuous stream of
a video (episode or movie) from the start of a video to the end. For
each user, the total number of watches of a TV show is the aggregation
of the number of times he/she has watched any episode of that show.

\begin{figure}
\centering
\includegraphics[width=3.5in]{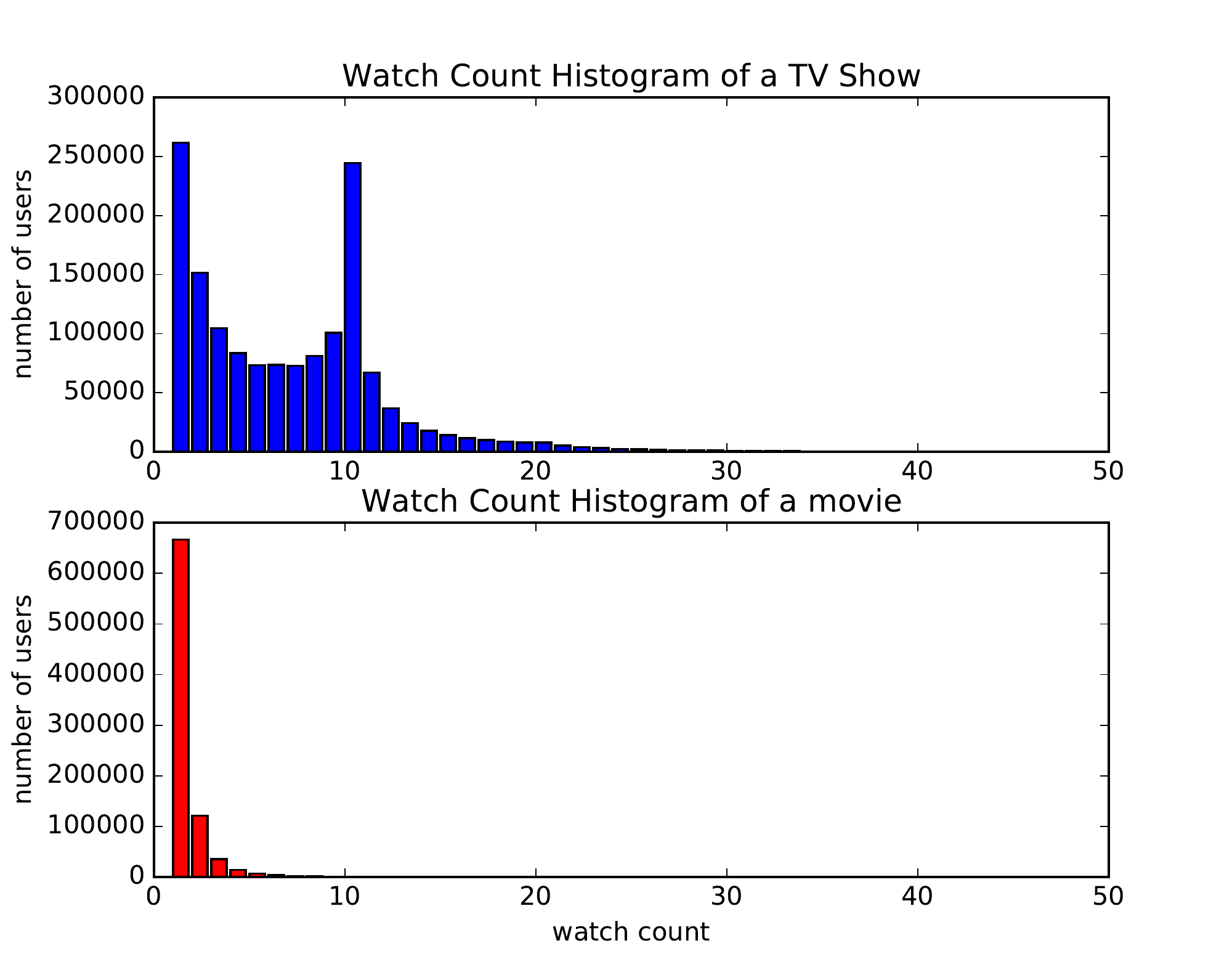}
\caption{The histogram of watch count for a TV show (above) and a movie (below).
The x-axis is the watch count  and the y-axis is the number of users who have watched the item.}
\label{fig:show_movie}
\end{figure}

As the number of episodes varies dramatically between different TV
shows, and a movie usually has only one video, there would be a strong
bias if we directly use the number of times for each TV shows and
movie. To amend this problem, we propose to use a relative score
$\tilde{r}^u_i$ as a surrogate of the ``rating'' $r^u_i$ in
Equation~\ref{eqn:t} and Equation~\ref{eqn:confidence}, which is the
percentage of users whose watch count of item $i$ is no larger than
user $u$'s out of all users who have watched item $i$. For example, if
a user watched a TV show $i$ $10$ times (counting all its episodes)
and $100$ out of $1000$ people watched this TV show more no more than
$10$ times, where $1000$ is the total number of people who have
watched it, then the relative ``rating'' $\tilde{r}^u_i$ will be
$0.1$.
This method also applies for movies, as we observed that there are
always users who watch a movie multiple
times. Figure~\ref{fig:show_movie} shows the histograms of watch count
for a TV show and a movie, respectively. We can see that there are
users who have watched a movie even more than $5$ times. If a user's
watch count on a specific item (a TV show or a movie) is the largest
among all users who have watched the item, we have a good reason to
assume that the user likes this item very much. Moreover, we can also
observe that most of users watch a movie only once while watch a TV
show more than 1 time (namely, more than one episode). Hence, the
relative ``rating'' of a movie watched once will be higher than a TV
show also watched once by a user. This is a desirable characteristic
as watching a movie entirely strongly indicates that the user likes
the movie, while this does not hold if a user has watched only one
episode of a TV show. Another advantage of adopting relative ``ratings''
is that they take values in $[0,1]$, easy to be interpreted as
confidence levels. Thus, we replace $r^u_i$ with $\tilde{r}^u_i$ in
Equations~\ref{eqn:t} and \ref{eqn:confidence}, to compute the
\textit{like} vector $\bf t$ and the confidence vector $\bf c$
throughout all experiments.


\subsection{Evaluation Metric}
\label{sec:metric}

The difficulty with evaluating collaborative filtering methods in the
implicit feedback scenario is that we can only test the performance of
the models on items that a user has watched before, since
non-engagement is noisy and does not necessarily mean negative
feedback.
As a result, we follow the famous work of \cite{hu2008collaborative}
and use a recall based evaluation metric \emph{mean percentage
  ranking} (MPR).

Specifically, for each user, we randomly select $10\%$ relative
``ratings'' $\tilde{r}^u_i$ on TV shows or movies he/she has watched, as
the test set. The samples in the test set are denotes as
$\tilde{\gamma}^u_i$ and their counterparts $\tilde{r}^u_i$ in the
training are set to $0$. Thus, the training set does not contain any
watch behaviors of the test samples, and the goal of the CF models is
to predict the users' preference on the test samples.  For each user,
we generated a ranked list of all candidate items sorted by preference
from Equation~\ref{eqn:test_cond}. Let $\textup{rank}^u_i$ be the
percentile ranking of item $i$ for user $u$ in the ranked list, where
$\textup{rank}^u_i=0\%$ means that $i$ is predicted as the highest
recommended item for $u$, and $\textup{rank}^u_i=100\%$ signifies
lowest.
The MPR is then defined as:
\begin{equation}
\textup{MPR}=\frac{\sum_{u,i}\gamma^u_i \textup{rank}^u_i}{\sum_{u,i}\gamma^u_i}.
\label{eqn:mpr}
\end{equation}

Lower values of MPR indicate that users will watch the TV shows or
movies higher in the ranked list, which is desirable in practice. As
an extreme example, a randomly shuffled list would have an expected
MPR of $50\%$.

\subsection{Results}
\label{sec:results}

\begin{figure}
\centering
\includegraphics[width=3.5in]{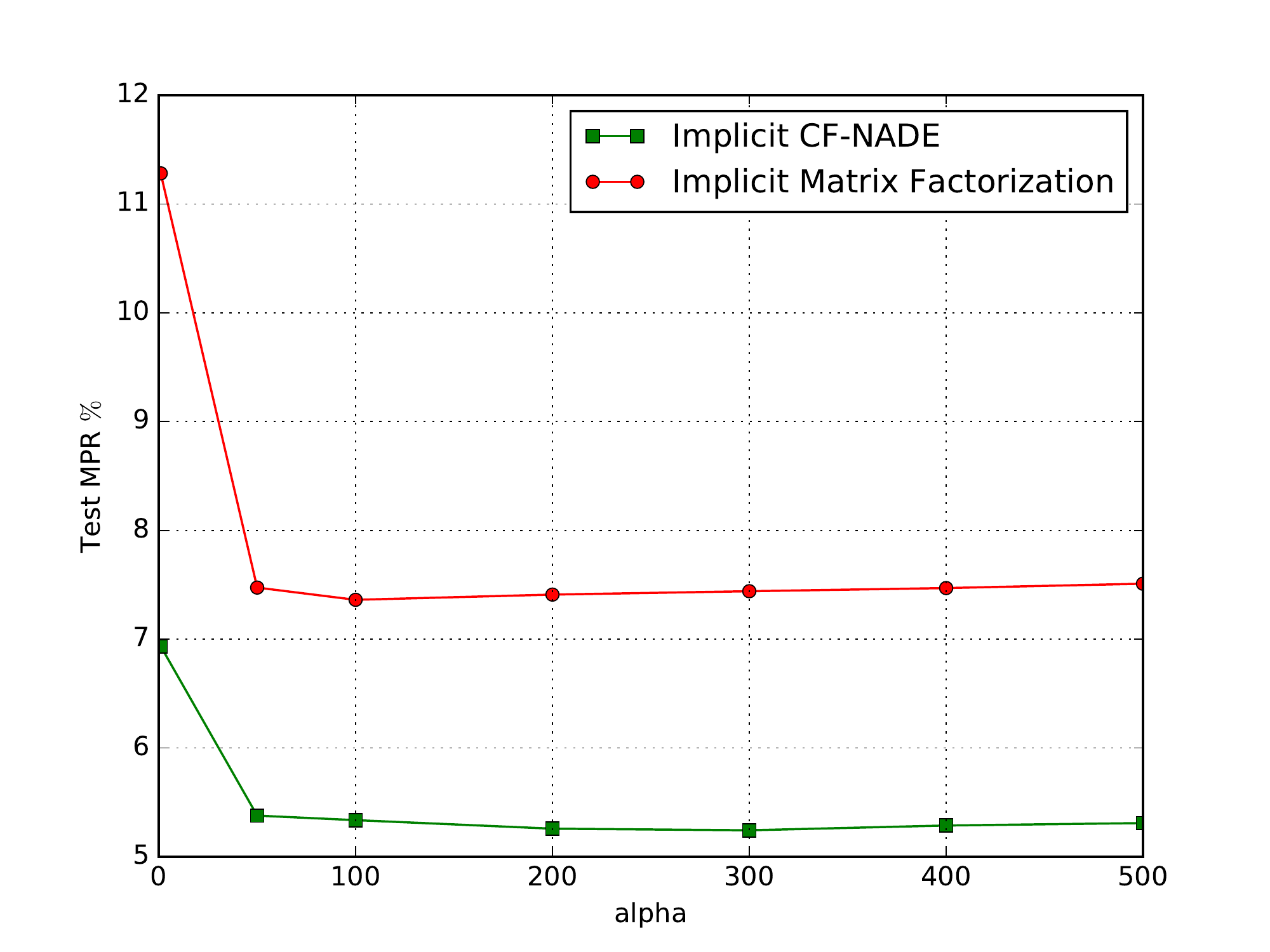}
\caption{Test MPRs for implicit \cfnade and implicit matrix
  factorization w.r.t $\alpha$ in Equation~\ref{eqn:confidence}.}
 \label{fig:mpr}
\end{figure}

In this section, we compare implicit \cfnade with Implicit Matrix
Factorization (IMF)~\cite{hu2008collaborative}. Implicit \cfnade is
implemented using Tensorflow~\cite{abadi2016tensorflow}, and IMF is
implemented using Spark MLlib~\cite{meng2015mllib}. In the
experiments, the implicit \cfnade model is trained with stochastic
gradient decent optimizer with learning rate set to $0.01$, batch size
$200$ and weight decay $0.01$. We use only one hidden layer and the
number of hidden units is set to $256$. The number of factors for IMF
is also set to $256$ for a fair comparison. Figure~\ref{fig:mpr}
depicts MPRs on the test set for different choices of the rate of
increase $\alpha$ in Equation~\ref{eqn:confidence}. We can see
that implicit \cfnade always outperforms IMF, and $\alpha$ should be
set to higher than $100$ for good performance for both algorithms.
The best test MPR ($5.2436\%$) of implicit \cfnade is achieved at
$\alpha=300$, and for IMF, the best test MPR is $7.362\%$ with
$\alpha=100$.

\section{Conclusion}
In this paper, we generalized the recently developed \cfnade to
\emph{implicit} \cfnade for real-world collaborative filtering tasks,
using implicit feedback. Specifically, we convert a user's watch
counts into implicit relative ratings, and then compute a ``like''
vector and a confidence vector. Implicit \cfnade is constructed by
modifying \cfnade to be aware of the ``like'' and confidence vector,
in the sense that the joint probability of a ``like'' vector
conditioned on the confidence vector is decomposed into conditionals
by chain rule, and the conditionals are modeled by a series of weight
sharing neural networks. We augmented the training loss with the
confidence vector, taking the uncertainty of ratings into
consideration. Experimental results show that implicit \cfnade
outperforms implicit matrix factorization on a dataset extracted from
a popular digital TV streaming service. This also indicates that
implicit \cfnade is highly effective in real-world applications.


%
\bibliographystyle{abbrv}
\bibliography{sigproc}  
\end{document}